\documentstyle[aps,pre,twocolumn]{revtex}
\begin{document}
\draft
 \def\OP{\tensor P}
\def\B.#1{{\bbox{#1}}}
\def\BE {\begin{equation}}
\def\EE {\end{equation}}
\def\BEA {\begin{eqnarray}}
\def\EEA {\end{eqnarray}}
%%%%%%%%%%%%%%%%%%%%%%%%%%%%%%%%%%%%%%%%%%%%%%%%%%%%%%%%%%%%%
\title{
 Hydrodynamic Turbulence Has Infinitely Many
Anomalous Dynamical Exponents}
  \author {Victor S. L'vov$^{1,2}$, Evgenii
Podivilov$^{1,2}$
and Itamar  Procaccia$^1$}
  \address{${^1}$Department of~~Chemical Physics,
 The Weizmann Institute of Science,
  Rehovot 76100, Israel,\\
  ${^2}$Institute of Automation and Electrometry,
   Ac. Sci.\  of Russia, 630090, Novosibirsk, Russia}
 \maketitle
\begin{abstract}
   On the basis of the Navier-Stokes equations we develop the
   statistical theory of many space-time correlation functions of
   velocity differences. Their time dependence is {\em not} scale
   invariant: $n$-order correlations functions exhibit $n-1$ distinct
   decorrelation times that are characterized by $n-1$ anomalous
   dynamical scaling exponents. We derive exact scaling relations that
   bridge all these dynamical exponents to the static anomalous
   exponents $\zeta_n$ of the standard structure functions.
\end{abstract}
%\begin{multicols}{2}
%%%%%%%%%%%%%%%%%%%%%%%%%%%%
Experimental investigations of the statistical objects that
characterize the small scale structure of turbulent flows are almost
invariably based on a single point measurement of the velocity field
as a function of time \cite{MY-2,Fri,94Nel,96AS}.  The Taylor ``frozen
turbulence" hypothesis is then used to surrogate time for space. The
results of this type of measurements are ``simultaneous" correlation
of the velocity field itself or of velocity differences across a scale
$R$ (structure functions), or of velocity gradient fields like the
dissipation field.  Theoretical analysis which starts with the
Navier-Stokes equations, on the other hand, states unequivocally that
a closed form theory for the simultaneous many space-point correlation
functions of velocity differences is not available. The perturbative
theory of turbulence is very clear about this: attempting to derive
equations for simultaneous correlation functions one finds integrals
over time differences of space-time correlation and response functions
\cite{P-1}. There are no small parameters like a ratio of time scales
(as in turbulent advection \cite{68Kra}) or a small interaction (like
in weak turbulence \cite{ZLF}) that allow a reduction of such a theory
to a closed scheme in terms of simultaneous objects only. The aim of
this Letter is to initiate a {\em nonperturbative} analytic theory of
$n$-order space-time correlation functions, and to find the
characteristics of such a theory that can be related to objects that
are known from standard experiments. It should be pointed out that in
addition to the fundamental interest of such a theory it has important
applications for the theory of scalar advection \cite{59Bat,74Kra};
such a theory is beyond the scope of this Letter and is only mentioned
as an additional motivation for the present analysis.

In recent theoretical work \cite{96LP,NP-1} about the same time
statistics of turbulence it was made clear why it is extremely
advantageous to consider many point correlation functions. The basic
field under study is the difference of the Eulerian velocity field
${\B.u}({\B.r},t)$ across a scale $\B.R\equiv {\B.r}'-{\B.r}$: $
{\B.w}({\B.r},{\B.r}',t) \equiv {\B.u} ({\B.r}',t)- {\B.u}({\B.r},t)
$.  The fundamental statistical quantities are the simultaneous
``fully unfused" $n$-rank tensor correlation function of velocity
differences:
\begin{eqnarray}
&&{\B.F}_n({\B.r}_1,{\B.r}'_1;{\B.r}_2,{\B.r}'_2;
\dots;{\B.r}_n,{\B.r}'_n)
\nonumber \\
&=& \left< \B.w({\B.r}_1,{\B.r}'_1,t)
\B.w({\B.r}_2,{\B.r}'_2,t) \dots
\B.w({\B.r}_n,{\B.r}'_n,t) \right> \ ,
\label{defF}
\end{eqnarray}
where pointed brackets denote the cumulant part of the average over a
(time-stationary) ensemble. In this quantity all the coordinates are
distinct. The more usual structure function $S_n(R)$
\begin{equation}
S_n(R) = \left\langle\vert\ {\B.w}({\B.r},{\B.r}',t)
\vert^n\right\rangle \ , \quad \B.R\equiv {\B.r}'-{\B.r}\ ,\label{a5}
\end{equation}
is obtained by fusing all the coordinates $\B.r_i$ into one point
$\B.r$, and all the coordinates $\B.r'_i$ into another point
$\B.r+\B.R$. Obviously, in using the functions of many variables
${\B.F}_n$ instead of the one variable function $S_n(R)$ one is paying
a heavy price. On the other hand this has an enormous advantage: when
one develops the theory for $S_n(R)$ on the basis of the Navier-Stokes
equations one encounters the notorious closure problem: knowing
$S_n(R)$ requires information about $S_{n+1}$ etc. It is well known
that arbitrary closures of this hierarchy of equations are doomed,
leading to predictions that are in contradiction with experiments. On
the other hand no one succeeded to solve the hierarchy in its
entirety. In contradistinction, the theory for the fully unfused
${\B.F}_n$ does not suffer from this problem: it was shown that there
exist homogenous equations for ${\B.F}_n$ in terms of ${\B.F}_n$,
without any hierarchic connections to higher or lower order
correlation functions. This fact allows one to proceed
\cite{96LP,NP-1} to derive a variety of exact bridge relations between
the scaling exponents of gradient fields and the scaling exponents of
the structure functions themselves, and to study the nature of the
dissipative scales in turbulence, showing that in fact they are
scaling functions with well defined scaling exponents.

In considering the decorrelation times of many ``fully unfused"
space-{\em time} correlation functions we need to make a choice of
which velocity field we take as our fundamental field.  The Eulerian
velocity field won't do, simply because its decorrelation time is
dominated by the sweeping of small scales by large scale flows. In
\cite{P-1} we showed that at least from the point of view of the
perturbative theory one can get rid of the sweeping effect using the
Belinicher-L'vov velocity fields whose decorrelation time is intrinsic
to the scale of consideration.  In terms of the Eulerian velocity
ref.\cite{87BL} defined the field ${\B.v}({\bf r}_0,t_0\vert
{\B.r},t)$ as
 \begin{eqnarray}
 {\B.v}({\B.r}_0,t_0\vert {\B.r},t)&\equiv& {\B.u}\lbrack{\B.r}
 +\B.\rho_{_{\rm L}}({\B.r}_0,t_0|t),t\rbrack \ ,
 \label{a2}\\
 \B.\rho_{_{\rm L}}({\B.r}_0,t_0|t)
 &=&\int_{t_0}^{t}{\B.u }[{\B.r}_0 +\B.\rho_{_{\rm L}}
(\B.r_0,t_0|\tau) ,\tau] \ .
  \label{a3}
 \end{eqnarray}
 The observations of Belinicher and L'vov was that the variables
 ${\B.v}({\B.r}_0,t_0\vert {\B.r},t)$ satisfy a Navier-Stokes-like
 equation in the limit of incompressible fluid, and that their
 simultaneous correlators are identical to the simultaneous
 correlators of $\B.u(\B.r,t)$.
 
 Introduce now a difference of two (simultaneous) BL-velocities at
 points $\B.r$ and $\B.r'$
\begin{equation}
\B.{\cal W}(\B.r_0,t_0|\B.r,\B.r',t)\equiv   \B.v( {\B.r}_0,t_0| {\B.r},t)-
\B.v( {\B.r}_0,t_0| {\B.r}',t) \ . \label{newBL}
\end{equation}
The equation of motion for $\B.{\cal W}$ can be
calculated starting from the Navier-Stokes equation for the Eulerian field,
\FL
\begin{equation}
\Big[{\partial\over \partial t}\!+\!\hat{\cal L}(\B.r,\B.r',t)
\!-\!\nu (\nabla_r^2+\nabla_r'^2)\Big]
\B.{\cal W}(\B.r_0,t_0|\B.r,\B.r',t)=0. \label{newNS}
\end{equation}
We introduced an operator $ \hat{\cal L}(\B.r_0,t_0|\B.r,\B.r',t)$
\begin{eqnarray}
 \hat{\cal L}(\B.r_0,t_0|\B.r,\B.r',t)&\equiv&
\tensor{\B.P} \B.{\cal W}({\B.r}_0,t_0|{\B.r},\B.r_0,t)\cdot{\B.\nabla_r}
\nonumber\\ &+&
\tensor{\B.P}' \B.{\cal W}({\B.r}_0,t_0|{\B.r'},\B.r_0,t)\cdot{\B.\nabla_r'}\ ,
\label{calL}
 \end{eqnarray}
 where $\tensor{\B.P}$ is the usual transverse projection operator
 which is formally written as $\tensor{\B.P}$ $\equiv
 -\nabla^{-2}{\bbox{\nabla }}\times\B.\nabla\times$. The application
 of ${\tensor{\B.P} }$ to any given vector field $ {\B.a}( {\B.r})$ is
 non local, and it has the form:
 \begin{equation}
 \lbrack\tensor{\B.P}{\B.a}(
 {\B.r})\rbrack_\alpha =\int d  \tilde\B.r P_{\alpha\beta}(
 {\B.r}-\tilde\B.r)a_\beta(\tilde\B.r).
 \label{b2}
 \end{equation}
 The explicit form of the kernel can be found, for example, in
 \cite{P-1}.  In (\ref{calL}) $\tensor{\B.P}$ and $\tensor{\B.P}'$ are
 projection operators which act on fields that depend on the
 corresponding coordinates $\B.r$ and $\B.r'$. The equation of motion
 (\ref{newNS}) form the basis of the following discussion of the time
 correlation functions.
 
 To simplify the appearance of the fully unfused, multi-time
 correlation function of BL-velocity differences we choose the
 economic notation $\B.{\cal W}_j\equiv \B.{\cal
   W}({\B.r}_0,t_0|{\B.r_j}\B.r'_j,t)$:
\begin{equation}
\B.{\cal F}_n({\B.r}_0,t_0|{\B.r_1}\B.r'_1,t_1 \dots {\B.r_n}\B.r'_n,t_n)
= \left< \B.{\cal W}_1 \dots \B.{\cal W}_n \right> \ .
\label{defFtime}
\end{equation}
We begin the development with the simplest non- simultaneous case in
which there are two different times in (\ref{defFtime}).  Chose
$t_i=t+\tau$ for every $i\le p$ and $t_i=t$ for every $i>p$. We will
denote the correlation function with this choice of times as $\B.{\cal
  F}_{n,1}^{(p)}(\tau)$, omitting for brevity the rest of the
arguments.  Compute the time derivative of $\B.{\cal F}_{n,1}^{(p)}$
with respect to $\tau$:
\begin{equation}
{\partial \B.{\cal F}_{n,1}^{(p)}(\tau)\over d\tau}
=\sum_{j=1}^p\langle \B.{\cal W}_1\dots
{\partial \B.{\cal W}_j\over \partial t}\dots
\B.{\cal W}_n \rangle \ , \label{dt}
\end{equation}
Using the equation of motion (\ref{newNS}) we find
\begin{eqnarray}
&&{\partial \B.{\cal F}^{(p)}_{n,1}(\tau)\over d\tau}
+\B.{\cal D}^{(p)}_{n,1}(\tau)=\B.{\cal J}^{(p)}_{n,1}(\tau)
\ , \label{balnp}\\
&&\B.{\cal D}^{(p)}_{n,1}(\tau) = \sum_{j=1}^p\langle \B.{\cal W}_1\dots
\hat\B.{\cal L}_j \B.{\cal W}_j\dots \B.{\cal W}_n \rangle \ ,
 \label{defDnp}\\
&&\B.{\cal J}^{(p)}_{n,1} (\tau)
= \nu\sum_{j=1}^p(\nabla_j^2+{\nabla'}_j^2)\langle
\B.{\cal W}_1\dots \B.{\cal W}_j\dots \B.{\cal W}_n \rangle \ ,
\label{defJnp}
\end{eqnarray}
with $\hat\B.{\cal L}_j\equiv\hat\B.{\cal
  L}(\B.r_0,t_0|\B.r_j,\B.r'_j,t)$.  We remember that $\hat\B.{\cal
  L}_j \B.{\cal W}_j$ is a nonlocal object that is quadratic in
BL-velocity differences, cf.  Eq.(\ref{calL}). We reiterate that all
the functions depend on $2n$ space coordinates that we do not display
for notational economy.

To understand the role of the various terms in Eq.(\ref{balnp}) we
will make use of the analysis of a similar equation that was presented
in \cite{NP-1}. In that case we considered the simultaneous object
$\B.F_n=\B.{\cal F}^{(p)}_{n,1}(0)$ and computed, as above, its
$t$-time derivative.  Obviously, in the stationary ensemble this
derivative vanishes.  Instead of $\B.{\cal D}^{(p)}_{n,1}(\tau)$ and
$\B.{\cal J}^{(p)}_{n,1}(\tau)$ we got $\B.{\cal D}_{n,1}^{(n)}(\tau)$
and $\B.{\cal J}_{n,1}^{(n)}(\tau)$ which differ from the present ones
only in that the summation go up to $n$ instead of $p$.  The crucial
observations of \cite{NP-1} are the following ones: (i) $\lim_{\nu\to
  o} \B.{\cal J}_{n,1}^{(n)}(\tau)=0$. For fully unfused simultaneous
correlation functions the viscous term in the balance equation
disappears in the limit of vanishing viscosity.  (ii) The integral in
$\B.{\cal D}_{n,1}^{(n)}(\tau)$ which originates from the projection
operator converges in the infra-red and the ultraviolet regimes. This
means that every term in the sum over $j$ that contributes to
$\B.{\cal D}_{n,1}^{(n)}(\tau)$ can be estimated as $S_{n+1}(R)/R$
when all the separation $\B.R_j\equiv \B.r_j-\B.r'_j$ are of the same
order $R$.  When we take the full sum up to $n$ there exist internal
cancellations between all these terms, leading to the homogeneous
equation $\B.{\cal D}_{n,1}^{(n)}=0$.

In the present analysis the proof that the viscous term $\B.{\cal
  J}^{(p)}_{n,1}(\tau)$ is negligible when $\nu\to 0$ is an immediate
consequence of the previous result. Since we are taking only partial
sums on $j$, the internal cancellation in$\B.{\cal
  D}^{(p)}_{n,1}(\tau)$ disappears, and it has a finite limit when
$\nu\to 0$. On the other hand $\B.{\cal J}^{(p)}_{n,1}(\tau)$ can only
increase if we take $\tau=0$.  Thus again when $\nu\to 0$ we can
neglect $\B.{\cal J}^{(p)}_{n,1}(\tau)$ in the fully unfused
situation. Accordingly for $\nu\to 0$ we have
\begin{equation}
\partial \B.{\cal F}^{(p)}_{n,1}(\tau)/ d\tau
+\B.{\cal D}^{(p)}_{n,1}(\tau)=0\ .
 \label{balt}
\end{equation}
The proof of convergence of the integrals in $\B.{\cal
  D}^{(p)}_{n,1}(\tau)$ follows from the previous results, since the
time correlation functions are bounded from above by the simultaneous
ones.  Accordingly, when all the coordinates are fully unfused and the
separations are all of the order of $R$,
\begin{equation}
\B.{\cal D}^{(p)}_{n,1}(\tau)\sim \B.{\cal
  F}_{n+1,1}^{(p+1)}(\tau)/R \ . \label{evalD}
\end{equation}
Introduce now the typical decorrelation time $\tau_{n,1}^{(p)}(R)$
that is associated with the one-time difference quantity $\B.{\cal
  F}^{(p)}_{n,1}(\tau)$ when all the separations are of the order of
$R$:
\begin{equation}
\int_{-\infty}^0 d\tau \B.{\cal F}^{(p)}_{n,1}(\tau) \equiv
\tau_{n,1}^{(p)} \B.{\cal F}^{(p)}_{n,1}(0) \ . \label{deftau1}
\end{equation}
Remember that the simultaneous correlation functions of BL-velocity
differences (\ref{defFtime}) are identical \cite{P-1} to the
simultaneous correlation functions of Eulerian velocity differences
(\ref{defF}), i.e.  $\B.{\cal F}^{(p)}_{n,1}(0)=\B.F_n$.  Integrate
(\ref{balt}) in the interval $(-\infty,0)$, use the evaluation
(\ref{evalD}), and derive
\begin{equation}
R \B.F_n \sim \tau_{n+1,1}^{(p+1)}\B.F_{n+1}\ .
\end{equation}
We see that from the point of view of scaling there is no $p$
dependence in this equation: for different values of $p$ only the
coefficients can change. We thus estimate
\begin{equation}
\tau_{n,1}(R) \sim R S_{n-1}(R)/ S_n(R)\propto  R^{z_{n,1}}\ . \label{tau1}
\end{equation}
Here we introduced the dynamical scaling
exponent $z_{n,1}$ that characterizes this time and found that
\begin{equation}
z_{n,1} = 1+\zeta_{n-1}-\zeta_n \ . \label{zn1}
\end{equation}

Consider next the three-time quantity that is obtained from $\B.{\cal
  F}_n$ by choosing $t_i=t+\tau_1$ for $i\le p$, $t_i=t+\tau_2$ for
$p<i\le p+q$, and $t_i=t$ for $i>p+q$. We denote this quantity as
$\B.{\cal F}_{n,2}^{(p,q)}(\tau_1,\tau_2)$, omitting again the rest of
the arguments.  We define the decorrelation time $\tau_{n,2}^{(p,q)}$
of this quantity by
\begin{equation}
\int_{-\infty}^0 d\tau_1 d\tau_2 \B.{\cal F}_{n,2}^{(p,q)}(\tau_1,\tau_2)
\equiv
[\tau_{n,2}^{(p,q)}]^2 \B.{\cal F}_{n,2}^{(p,q)}(0,0)\ . \label{deftau2}
\end{equation}
One could think naively that the decorrelation time
$\tau_{n,2}^{(p,q)}$ is of the same order as (\ref{tau1}). The
calculation leads to a different result. To see this calculate the
double derivative of $\B.{\cal F}_{n,2}^{(p,q)}(\tau_1,\tau_2)$ with
respect to $\tau_1$ and $\tau_2$. This results in a new balance
equation.  For the fully unfused situation, and in the limit $\nu\to
0$ we find
\begin{equation}
\partial^2 \B.{\cal F}_{n,2}^{(p,q)}(\tau_1,\tau_2)
/ \partial \tau_1 \partial \tau_2+\B.{\cal D}_{n,2}^{(p,q)}(\tau_1,\tau_2)=0
\ , \label{bal2t}
\end{equation}
where now
$$
\B.{\cal D}_{n,2}^{(p,q)}(\tau_1,\tau_2)
\!=\!\sum_{j=1}^p\!\sum_{k=p+1}^{p+q}\!\langle \B.{\cal W}_1\!\dots\!
\hat\B.{\cal L}_j \B.{\cal W}_j\!\dots\! \hat\B.{\cal L}_k \B.{\cal
W}_k\!\dots\!
\B.{\cal W}_n \rangle.
$$
In the RHS of (\ref{bal2t}) we neglected two terms that vanish in
the limit $\nu \to 0$.  The expression for $\B.{\cal
  D}_{n,2}^{(p,q)}(\tau_1,\tau_2)$ contains two space-integrals that
originate from the two projection operators which are hidden in
$\hat\B.{\cal L}_j$ and $\hat\B.{\cal L}_k$. Using the same ideas when
all the separations are of the order of $R$ we can estimate with
impunity
\begin{equation}
\B.{\cal D}_{n,2}^{(p,q)}(\tau_1,\tau_2)
\sim \B.{\cal F}_{n+2,2}^{(p+1,q+1)}(\tau_1,\tau_2)/R^2 \ . \label{est2}
\end{equation}
Integrating now Eq.(\ref{bal2t}) over $\tau_1$ and $\tau_2$ in the
interval $[-\infty,0]$ and remembering that $\B.{\cal
  F}_{n,2}^{(p,q)}(0,0)= \B.F_n$ we find
\begin{equation}
R \B.F_n \sim \B.F_{n+2} [\tau_{n+2,2}^{(p+1,q+1)}]^2\ .
\end{equation}
As before the scaling exponents are independent of $p$ and $q$ and we
we can introduce a notation $\tau_{n,2}\sim\tau_{n,2}^{(p,q)}$. Up to
$p,q$-dependent coefficients
\begin{equation}
[\tau_{n,2}(R)]^2 \sim R^2 S_{n-2}(R)/ S_n(R)\propto [R^{z_{n,2}}]^2
\ . \label{tau2}
\end{equation}
We see that the naive expectation is not realized. The scaling
exponent of the present time is different from (\ref{zn1}):
\begin{equation}
z_{n,2} =1+(\zeta_{n-2}-\zeta_n)/2 \ . \label{zn2}
\end{equation}
The difference between the two scaling exponents $z_{n,1}-z_{n,2}=
\zeta_{n-1}-(\zeta_n+\zeta_{n-2})/2$. This difference is zero for linear
scaling,
meaning that
in that case the naive expectation that the time scales are
identical is correct. On the other hand for the situation of
multiscaling the Hoelder inequalities require the difference to be
positive. Accordingly, for $R\ll L$ we have $\tau_{n,2}(R)\gg \tau_{n,1}(R)$.

We can proceed with correlation functions that depend on $m$ time
differences. Omitting the upper indices which are irrelevant
for the scaling exponents we denote the correlation function as
$\B.{\cal F}_{n,m}(\tau_1\dots \tau_m)$, and establish the
exact scaling law for its decorrelation time.
The definition of the decorrelation time is
\begin{equation}
\int_{-\infty}^0 d\tau_1\dots d\tau_m\B.{\cal F}_{n,m}(\tau_1\dots \tau_m)
\equiv [\tau_{n,m}]^m \B.{\cal F}_{n,m}(0\dots 0).\label{gentau}
\end{equation}
Repeating the steps described above we find the dynamical scaling
exponent that characterizes $\tau_{n,m}$ when all the separations are
of the order of $R$, $\tau_{n.m}\propto R^{z_{n,m}}$:
\begin{equation}
z_{n,m} =1 +(\zeta_{n-m}-\zeta_n)/m \ , \quad n-m\le 2 . \label{znm}
\end{equation}
One can see, using the Hoelder inequalities, that $z_{n,m}$ is a
nonincreasing function of $m$ for fixed $n$, and in a multiscaling
situation they are decreasing. The meaning is that the larger $m$ is
the {\em longer} is the decorrelation time of the corresponding
$m+1$-time correlation function, $
\tau_{n,p}(R)\gg \tau_{n,q}(R)\ \ {\rm for}\ \ p<q\ .
$

To gain further understanding of the properties of the time
correlation functions, we consider higher order temporal moments of
the two-time correlation functions:
\begin{equation}
\int_{-\infty}^0 d\tau \tau^{k-1} \B.{\cal F}^{(p)}_{n,1}(\tau) \equiv
({\overline{\tau^k}})^{(p)}_{n,1} \B.{\cal F}^{(p)}_{n,1}(0)\ .
\label{deftauk}
\end{equation}
The intuitive meaning of $({\overline{\tau^k}})^{(p)}_{n,1}$ is a {\em
  $k$-order decorrelation moment} of $\B.{\cal
  F}^{(p)}_{n,1}(R,\tau)$, and note that its dimension is (time)$^k$.
The first order decorrelation moment is the previously defined
decorrelation time $\tau_{n,1}^{(p)}$. To find the scaling exponents
of these quantities we start with Eq.(\ref{balt}), multiply by
$\tau^k$, and integrate over $\tau$ in the interval $(-\infty,0)$.
Using the evaluation (\ref{evalD}) and assuming convergence of the
integrals over $\tau$ we derive
\begin{equation}
-k\int_{-\infty}^0 \B.{\cal F}^{(p)}_{n,1}\tau^{k-1} d\tau\sim
{1\over R}\int_{-\infty}^0 \B.{\cal F}^{(p+1)}_{n+1,1}\tau^k d\tau \ ,
\end{equation}
where we have integrated by parts on the LHS. We stress that in
deriving this equation we assert that $k+1$ moments exist; this is not
known apriori.  Using the definition (\ref{deftauk}) and for all the
separations of the order of $R$ we find the recurrence relation
\begin{equation}
RS_n(R)({\overline{\tau^k}})^{(p)}_{n,1}\sim S_{n+1}(R)
({\overline{\tau^k}})^{(p+1)}_{n+1,1}\ .
\end{equation}
The solution is
\begin{equation}
 [\overline{\tau^k}]_{n,1}^{(p)}\sim  (\tau_{n,k})^k \sim
{R^k S_{n-k}(R)\over  S_n(R)}\ . \label{relk}
\end{equation}
for $k \le n-2 $. The procedure does not yield information about
higher $k$ values.

We learn from the analysis of the moments that there is no single
typical time which characterizes the $\tau$ dependence of $\B.{\cal
  F}^{(p)}_{n,1}$.  There is no simple ``dynamical scaling exponent"
$z$ that can be used to collapse the time dependence in the form
$\B.{\cal F}^{(p)}_{n,1}(\tau)\sim R^{\zeta_n}f(\tau/R^z)$.  Even the
two-time correlation function is not a scale invariant object. In this
respect it is similar to the probability distribution function of the
velocity differences across a scale $R$, for which the spectrum of
$\zeta_n$ is a reflection of the lack of scale invariance.

The main conclusion of this Letter is that all the dynamical scaling
exponents can be determined from the knowledge of the scaling
exponents $\zeta_n$ of the standard structure functions $S_n(R)$. All
the scaling relations that were obtained above can be easily
remembered using the following simple rule.

To get the dynamical scaling exponent every integral over $\tau$ in
the definition of the decorrelation time (\ref{gentau}), and every
factor $\tau$ in the definition of the moments (\ref{deftauk}) can be
traded for a factor of $R/{\cal W}$ {\em within the average} of the
correlation function involved. The dynamical exponent is determined by
the resulting scaling exponents of the resulting simultaneous
correlation function.

The deep reason for this simple rule is the non-perturbative locality
(convergence) of the integrals appearing in ${\cal D}_n$. Because of
this locality one can estimate from the equation of motion
(\ref{newNS}) $ {1/ \tau}\sim {\partial/\partial \tau}\sim \B.{\cal
  W}\cdot\B.\nabla \sim {{\cal W}(R)/R} $. This means that we can use
the substitutions
\begin{equation}
\tau,\int d\tau \Rightarrow R/{\cal W}(R)
\end{equation}
{\em as long as we use them within the average}, and when all the
separations are of the order of $R$. We propose to refer to this
substitution rule as ``weak dynamical similarity", where ``weak"
stands for a reminder that the rules can be used {\em only} under the
averaging procedure, and {\em only} for scaling purposes. The same
property of locality of the interaction integrals was shown
\cite{96LP,NP-1} to yield another set of bridge relations between
scaling exponents of correlation functions of gradient fields and the
scaling exponents $\zeta_n$. Those relations can be summarized by
another substitution rule that we refer to as ``weak dissipative
similarity"; It follows from equating the viscous and nonlinear terms
in the equations of motion:
\begin{equation}
\nu \nabla^2 \Rightarrow {\cal W}(R)/R.
\end{equation}
Again ``weak" refers to the reminder that we are only allowed to use
these substitutions for scaling purposes within the average. Note that
our weak dissipative similarity rule is weaker than the Kolmogorov
refined similarity hypothesis which states the {\em dynamical}
relationship $(\nabla {\cal W}(R))^2\sim {\cal W}^3/R$. Both our rules
are derived from first principles, while Kolmogorov's hypothesis is a
guess.

Finally, we should ask whether the results presented above are
particular to the time correlation function of BL-velocity
differences, or do they reflect intrinsic scaling properties that are
shared by other dynamical presentations like the standard Lagrangian
velocity fields. The answer is that the results are general; all that
we have used are the property of convergence of the interaction
integrals, and the fact that the simultaneous correlation functions of
the BL-fields are the same as those of the Eulerian velocities. These
properties hold also for Lagrangian velocities, and in fact for any
sensible choice of velocity representation in which the sweeping
effect is eliminated.  Accordingly we state that the dynamical
exponent are invariant to the representation and in particular will be
the same for many-time correlation functions of Lagrangian velocity
differences.

%%%%%%%%%%%%%%%

This work was supported in part by the German Israeli Foundation, the
US-Israel Bi-National Science Foundation, the Minerva Center for
Nonlinear Physics, and the Naftali and Anna Backenroth-Bronicki Fund
for Research in Chaos and Complexity.
%%%%%%%%%%%%%%%%%

%\end{multicols}

\begin{references}

\bibitem{MY-2} A.~S. Monin and A.~M. Yaglom.  \newblock {\em
    Statistical Fluid Mechanics: Mechanics of Turbulence}, volume~II.
  \newblock (MIT Press, Cambridge, Mass., 1973).

\bibitem{Fri} Uriel Frisch.  \newblock {\it Turbulence: The Legacy of
    A.N. Kolmogorov}.  \newblock (Cambridge University Press,
  Cambridge, 1995).

\bibitem{94Nel}
M.~Nelkin.
\newblock {\em Advan. in Phys.}, {\bf 43},143 (1994).

\bibitem{96AS} K.R. Sreenivasan and R.A. Antonia, ``The Phenomenology
  of Small-Scale Turbulence", preprint 1996.
  
\bibitem{P-1} V.S. L'vov and I. Procaccia, Phys. Rev. E, {\bf 52},
  3840 (1995); {\bf 52}, 3858 (1995); {\bf 53},3468 (1996).
  
\bibitem{68Kra} R.H. Kraichnan, Phys. Fluids {\bf 11},945 (1968).
  
\bibitem{ZLF} V.E. Zakharov, V.S. L'vov and G. Falkovich, {\em
    Kolmogorov Spectra of Turbulence I. Weak Wave Turbulence}
  (Springer, Heidelberg 1992)
  
\bibitem{59Bat} G.K. Batchelor, J. Fluid Mech. {\bf 5}, 113 (1959)
  
\bibitem{74Kra} R.H. Kraichnan, J. Fluid Mech. {\bf 64}, 737 (1974)
  
\bibitem{96LP} V.S. L'vov and I. Procaccia, ``The viscous scales in
  hydrodynamic turbulence are anomalous scaling functions", Phys. Rev.
  Lett., submitted.  (chao-dyn@xyz.lanl.gov \# 9606018)
  
\bibitem{NP-1} V.S. L'vov and I. Procaccia ``Towards a Nonperturbative
  Theory of Hydrodynamic Turbulence: Fusion Rules, Exact Bridge
  Relations and Anomalous Viscous Scaling Functions", Phys.Rev. E,
  submitted. (chao-dyn@xyz.lanl.gov \# 9607006)
  
\bibitem{87BL} V.~I. Belinicher and V.~S. L'vov, Zh. Eksp. Teor. Fiz.
  {\bf 93}, 1269 (1987) [{\em Sov.~Phys.~JETP}, 66:303--313, 1987].

%%%%%%%%%%%%%%%%%%%%%%%%%%%%%%%%%

\end{references}
\end{document}